\documentclass[twocolumn,aps,floatfix,superscriptaddress,longbibliography]{revtex4-2}
\usepackage{amsmath,amssymb,eucal,graphicx,float,epstopdf,xparse,color}
\usepackage{epsfig,subfigure}

\usepackage[colorlinks=true, urlcolor=blue, anchorcolor=blue, citecolor=blue,filecolor=blue,linkcolor=blue,menucolor=blue]{hyperref}
\newcommand{\ii}{{\rm i}}

\newcommand{\Hil}{\mathcal{H}}

\def\Xint#1{\mathchoice
	{\XXint\displaystyle\textstyle{#1}}%
	{\XXint\textstyle\scriptstyle{#1}}%
	{\XXint\scriptstyle\scriptscriptstyle{#1}}%
	{\XXint\scriptscriptstyle\scriptscriptstyle{#1}}%
	\!\int}
\def\XXint#1#2#3{{\setbox0=\hbox{$#1{#2#3}{\int}$}
		\vcenter{\hbox{$#2#3$}}\kern-.5\wd0}}
\def\dashint{\Xint-}

\definecolor{ca}{rgb}{0.9,0,0}    

\definecolor{cb}{rgb}{0,0.5,0}    

\definecolor{cc}{rgb}{0,0,1}    

\begin{document}
\title{Current fluctuations in the Dyson Gas}

\author{Rahul Dandekar}
\affiliation{Institut de Physique Th\'{e}orique, CEA/Saclay, F-91191 Gif-sur-Yvette Cedex, France}
\author{P. L. Krapivsky}
\affiliation{Department of Physics, Boston University, Boston, Massachusetts 02215, USA}
\affiliation{Santa Fe Institute, Santa Fe, New Mexico 87501, USA}
\author{Kirone Mallick}
\affiliation{Institut de Physique Th\'{e}orique, CEA/Saclay, F-91191 Gif-sur-Yvette Cedex, France}

\begin{abstract}
We study large fluctuations of the current in a Dyson gas, a 1D system of particles interacting through a logarithmic potential and subjected to random noise. We adapt the macroscopic fluctuation theory to the Dyson gas and derive two coupled partial differential equations describing the evolution of the density and momentum. These equations are nonlinear and non-local, and the `boundary' conditions are mixed: some at the initial time and others at the final time. If the initial condition can fluctuate (annealed setting), this boundary-value problem is tractable. We compute the cumulant generating function encoding all the cumulants of the current. 
\end{abstract}
\maketitle

\section{Introduction}

Systems with long-range interactions, e.g., gravitational or Coulomb forces, exhibit non-extensive behaviors and other unusual thermodynamic features. The study of the collective properties of such systems has been and remains an active research area, see \cite{Chandra-43,Dyson69,Lieb,Ruffo09,Ruffo14} and references therein. For systems composed of many particles, a hydrodynamic description captures the average evolution \cite{Giacomin}. The hydrodynamic description ignores fluctuations caused by thermal noise, which may lead to rare events. These phenomena are amplified when the particles are confined to a line or a plane; low dimensional systems are prone to exhibit anomalous fluctuations due to geometric constraints, non-crossing conditions, or long-range correlations \cite{BouchaudGeorges}. Long-range correlations can be intrinsic or caused by dynamics---a system of particles with localized interactions and non-zero current can display long-range order and undergo phase transitions even in one dimension \cite{spohnTracer1990,Spohn91}. 

The correspondence between non-equilibrium phenomena and systems with long-range forces can be subtle. Long-range interactions can be dynamically generated by conditioning the evolution. For example, when non-interacting  Brownian particles on a line are conditioned not to cross between an initial time $t=0$ and a final time $t_f$,  the resulting system is equivalent to particles interacting through a logarithmic potential, i.e., a $1/r$-force \cite{grabinerBrownianl1999,gautieNoncrossing2019}. This model is identical to the Dyson Brownian motion that describes the dynamics of the eigenvalues of a (symmetric, hermitian, or symplectic) matrix whose elements are independent Brownian processes \cite{Dyson62, Forrester, Vivo18, pottersFirst2020}.

More generally, one can consider Brownian particles on an infinite line that interact through the power law potential, $V(x,y) = g s^{-1} | x - y |^{-s}$, a Riesz gas of exponent (or index) $s$. The study of equilibrium properties of Riesz gases has gained a renewed interest both in physics  \cite{Marcos17,Kundu19,Kundu21,Kundu22a,Kundu24} and mathematics \cite{Serfaty17,Serfaty17a,Serfaty18,Petrache20,Lewin22,Menon23,Serfaty24} communities. In a recent work \cite{RahulRiesz}, we considered dynamical fluctuations in the Riesz gas, such as the variance of the position of a tagged particle and of the time-integrated current through a given location (since particles do not cross, these two quantities are related \cite{Varadhan2013}).  Our approach is inspired by the macroscopic fluctuation theory (MFT) developed to study large deviations in systems far from equilibrium \cite{Bertini01,Bertini02,Derrida07,MFT15}.  If $s>1$, the interaction is effectively local. Such Riesz gases behave as a single-file \cite{Dipoles-1d,KRB,Diamant,Barkai13}, the typical displacement of a tagged particle scales as $t^\nu$ with $\nu=\frac{1}{4}$, i.e., slower than the $\sqrt{t}$ diffusive scaling; the exponent $\nu$ is universal (independent on the exponent $s>1$). When $0 < s <1$, non-local effects prevail, and the spreading of the tagged particle is even slower with non-universal exponent $\nu=\frac{s}{2(s+1)}$.  In the $s \to 0$ limit, the interaction potential becomes logarithmic, the force between the particles decays as the inverse of the separation, and the Riesz gas becomes the Dyson Brownian gas. This limit is singular, and perturbation techniques \cite{Meerson12a,Kirone14} are ineffective at $s =0$. The analysis of the Dyson gas requires a different approach. 

In a series of pioneering articles \cite{Spohn87,Spohn87b,Spohn98MPRF},  H. Spohn showed that the variance of the position of a tagged particle in the Dyson gas grows logarithmically with time. We re-derive this anomalously slow growth law using the MFT. The MFT equations for the Dyson gas with long-range interactions differ from the standard MFT equations. We thus start with fluctuating hydrodynamics, more precisely the Dean-Kawasaki equation, and derive the MFT equations for the Dyson gas. The current and displacement of the tagged particle are ultimately related in one-dimensional systems in which crossing of the particles is impossible \cite{Varadhan2013}. We focus on the current and use the MFT equations to derive the cumulant generating function (CGF) encoding the variance and all higher cumulants of the current. Our computations rely on powerful analytical techniques developed in random matrix theory and complex analysis to study the Dyson Brownian motion \cite{Mehta,Forrester,Bouchaud,pottersFirst2020}.    

The outline of this work is as follows. In Sec.~\ref{sec:def}, we recall the dynamical equations of a Dyson gas, define the empirical density and the time-integrated current, and perform an elementary scaling analysis. We also introduce the CGF of the current, $\mu(\lambda)$, the fundamental quantity we shall study. Section \ref{sec:hydro} is devoted to the continuous description, in terms of fluctuating hydrodynamics:   in the spirit of the macroscopic fluctuation theory, we express the current CGF as a path integral and derive a pair of coupled non-linear partial differential equations that determine the saddle-point in the long time limit. We show that these equations are equivalent to Euler-type equations derived by Matytsin \cite{Matytsin}. We then map these equations into the complex Burgers equation. The analytical study of this equation is the subject of Sec.~\ref{sec:calculCGF}: a powerful symmetry of the problem allows us to find the optimal density profile at the final time, which yields the CGF  and all cumulants of the current. We conclude in Sec.~\ref{sec:concl}. We relegate to appendices several technical derivations.

\section{Dyson Gas}
\label{sec:def}

We consider a system of $N$  particles in one dimension. The particles are in a shallow external quadratic potential, they are subjected to a random noise and interact via a logarithmic potential. In the over-damped limit, the  positions $x_i$, $i=1,\ldots,N$, of the particles evolve according to stochastic differential equations \cite{Spohn87,Spohn87b,Spohn98MPRF}
\begin{equation}
  \frac{dx_i}{d t}
  =  g \sum_{j \neq i} \frac{1}{x_i-x_j} -  \epsilon x_i
  +  \sqrt{D}\, \eta_i(t) 
\label{Dyson0}
\end{equation}
Gaussian white noises $\eta_i(t)$ are independent, centered, identical and normalized:
\begin{equation}
\langle \eta_i(t)\rangle = 0, \qquad  \langle \eta_i(t_1) \eta_j(t_2) \rangle  = 2 \delta_{ij} \delta(t_1-t_2).
 \label{WhiteNoise}
\end{equation}
The shallow quadratic potential of small strength $\epsilon$ ensures that, at equilibrium,  the gas remains within $\pm\sqrt{2Ng/\epsilon}$. This property results from Wigner's semi-circle law since the stationary measure of Eq.~(\ref{Dyson0}) can be identified with the distribution of the eigenvalues of random matrices \cite{Mehta, Forrester, grabinerBrownianl1999, Vivo18,pottersFirst2020}. The interaction strength $g$ and the amplitude of the noise $D$ have the same dimension $[\text{length}]^2/[\text{time}]$. Their relative strength is quantified by the (dimensionless) ratio
\begin{equation}
  \beta =  \frac{g}{D} 
  \label{defbeta}
\end{equation}
playing the role of an inverse temperature. Exceptional values $\beta=1,2,4$ correspond to canonical ensembles (symmetric, hermitian, symplectic) of random matrices. For the Dyson gas, the ratio $\beta$ can be an arbitrary positive number ($\beta>0$ since the interactions are repulsive). The equilibrium characteristics of the model with arbitrary $\beta>0$ have been studied e.g. in \cite{dean06,majumdarHow2011,marinoNumber2016}. 

Taking the limits $N \to \infty$ and $\epsilon \to 0$ with $N \epsilon=\text{fixed}$ ensures that the particle density is essentially uniform and given by  
\begin{equation}
   \rho_0 = \frac{1}{\pi}\sqrt{ \frac{2 N \epsilon}{g}} \,. 
   \label{defrho0}
\end{equation}
In the large $N$ limit, we describe this ensemble of interacting  particles as a continuous fluid with local density
\begin{equation}
   \rho(x,t) = \sum_{i=1}^N \delta(x - x_i(t))   \,. 
\end{equation}
This stochastic density functional varies around the average density $\rho_0$. In this work, we are interested in fluctuations that occur in the Dyson gas evolving from the initial time $t =0$ until the final time $t_f$. We assume that the final time $t_f$ greatly exceeds the microscopic mean collision time $(g\rho_0^2)^{-1}$ but remains very small compared to the damping time  $\epsilon^{-1}$: 
\begin{equation}
   (g\rho_0^2)^{-1}    \ll    t_f \ll \frac{N}{g\rho_0^2}
\label{rangetf}
\end{equation}
Rescaling space and time variables 
\begin{eqnarray} 
 x_i'   =   \frac{x_i}{g\rho_0 t_f}  \,\,\,\,  {\rm  and }  \,\,\,\, 
 t'  = \frac{t}{t_f} \, ,
\end{eqnarray}
and neglecting the confinement term we recast \eqref{Dyson0} into
\begin{equation}
  \frac{ dx_i'}{d t'}  = \frac{1}{T} \sum_{j \neq i} \frac{1}{x_i'-x_j'}  
  +  \frac{1}{\sqrt{\beta T}}\,\eta_i(t') \, ,
  \label{Dyson1}
\end{equation}
where $\eta_i$'s  are normalized white noises and
\begin{equation}
    T =  g\rho_0^2 t_f \, .
\label{defT}
\end{equation}
The  conditions (\ref{rangetf}) imply that $1 \ll T \ll N$.
 
We define a normalized empirical density $q(x',t')$ as
\begin{equation}
  q(x',t') =   \frac{1}{T} \sum_{i=1}^N \delta(x_i' - x_i'(t'))
   = \frac{\rho(x,t)}{\rho_0}\, .
  \label{defq}
\end{equation}
Note that the typical value of $q(x',t')$ is 1. The time-integrated current $J$ through the origin from  the initial time $t =0$ up to  the final observation time $t_f$, is given by, taking  into account particle conservation,
\begin{eqnarray}
   J &=& \int_0^\infty dx\,[\rho(x,t_f) - \rho(x,0)] \nonumber  \\
    &=&   T  \int_0^\infty dx'\,[q(x',1) - q(x',0)] \,. 
   \label{TotCurr}
\end{eqnarray}
Our goal is to investigate the large deviation of the  random current $J$ in the $T \to \infty$ limit, or,  equivalently, its CGF
\begin{equation}
K(\gamma) = \log \langle {\rm e}^{\gamma J} \rangle = 
 \log \langle {\rm e}^{\lambda \beta T^2 \int_0^\infty dx\,[q(x',1) - q(x',0)]} \rangle \,.
  \label{CGF}
\end{equation}
For reasons that will become clear in the next section,  we have rewritten the  fugacity parameter $\gamma$  as 
\begin{equation}
   \gamma = \lambda\beta T \, . 
   \label{RescalLambda}
\end{equation}
The average in Eq.~\eqref{CGF}  depends on the stochastic evolution of the density and the choice of the initial density profile $q(x,0)$. The initial profile can be deterministic (by setting $q = 1$ at $t=0$) or drawn randomly from the equilibrium ensemble \cite{Varadhan2013,Meerson12a,Kirone14,DG09,Kurchan07}. In the latter {\it annealed} case, the equilibrium fluctuations of the initial profile are governed by the free energy \cite{Derrida07,MFT15,Kurchan07}. We limit ourselves to annealed initial conditions. The large deviations of the current $J$ will be obtained, in the long time limit, by taking the Legendre transform of the CGF.  Thus, at the dominant exponential order, we have   
\begin{equation}
     \log\left({\rm Prob}(J)\right) =
     \max_\gamma \left(  K(\gamma) - \gamma J \right) \, .
    \label{Legendre1}
 \end{equation}

\section{Continuous description of the Dyson Gas}
\label{sec:hydro}

\subsection{Fluctuating hydrodynamics for the density}

Following the procedure developed by Kawasaki \cite{kawasaki1994stochastic} and Dean \cite{dean96} we rely on fluctuating hydrodynamics, viz. a Langevin equation 
\begin{equation}
   \partial_t q  = \partial_x \left(
   \frac{C_\beta}{T}  \partial_x q -\pi q  \Hil[q ] +
   \frac{1}{T}\sqrt{\frac{q  }{\beta}} \, \eta(x,t) \right) 
   \label{DeanKawasaki}
\end{equation}
for the stochastic density $q(x,t)$. To avoid cluttering of formulas, hereinafter we use notation $x$ and $t$ for the dimensionless variables $x'$ and $t'$.

The white noise $\eta(x,t)$ satisfies 
\begin{equation}
   \langle \eta(x,t) \eta(x',t') \rangle  = 2 \delta(x - x') \delta(t-t') \,.
 \label{STWhite}
\end{equation}
The symbol $\Hil$  denotes the Hilbert transform, 
\begin{equation}
   \Hil[q](x,t) = \frac{1}{\pi}\dashint_{-\infty}^\infty dy\, \frac{q(y,t)}{x-y}\,,
   \label{def:Hilb}
 \end{equation}
where $\dashint$ denotes the Cauchy principal value 
\begin{equation*}
\dashint_{-\infty}^\infty dy\, \frac{q(y,t)}{x-y}=\lim_{\delta \to 0}\int_{|x-y|> \delta}
   dy\, \frac{q(y,t)}{x-y}\,.
\end{equation*}

The constant appearing in \eqref{DeanKawasaki} is $C_\beta = (2 -\beta)/(2\beta)$, see \cite{ChanPTRF,CepaLepingle,Touzo23}. In the $T \to \infty$ limit, the precise value of $C_\beta$ is irrelevant in our analysis. In Appendix~\ref{app:DKequation}, we recall the derivation of the Dean-Kawasaki equation \eqref{DeanKawasaki}.

The solution of the stochastic hydrodynamics equation \eqref{DeanKawasaki} can be represented as a path integral. This method was originally developed in  Refs.~\cite{MSR,Cyrano,Janssen} is now a standard procedure, see, e.g., \cite{DG09, Kirone15a,RahulRiesz} for recent applications to interacting particle systems. The bulk action in this path integral is given by $-\beta T^2  S[q,p]$ with  
\begin{equation}
   S[q,p] = p \partial_t q - q (\partial_x p)^2 -
   \pi q  \Hil[q] \partial_x p + \frac{C_\beta}{T} \partial_x p\, \partial_x q \, .
   \label{Action}
 \end{equation}  
The canonically conjugate fields  $q(x,t)$ and $p(x,t)$ represent the local density and momentum of the gas.

The stationary state of  Eq.~\eqref{DeanKawasaki} is an equilibrium measure,  $\frac{1}{Z} \exp(-  T^2 \beta {\mathcal F}[q])$,  where the free energy is given by 
\begin{eqnarray}
    {\mathcal F}[q] &=&
 - \frac{1}{2} \int \int dx dy\,\,{[q(x)-1][q(y)-1]}\,\log|x-y|\nonumber \\
  &+&  \frac{C_\beta}{T} \int dx\, q(x)\, \log{q(x)} \,.
  \label{FreeEnergy}
 \end{eqnarray}

The full action in the path integral that represents the CGF \eqref{CGF} is composed of the bulk action \eqref{Action}, the free energy \eqref{FreeEnergy}, and the time-integrated current \eqref{TotCurr}. The fact that the fugacity parameter has been rescaled as in Eq.~(\ref{RescalLambda})  ensures that these three terms appear with the same prefactor $\beta T^2$ (see Appendix~\ref{app:MFTpq}). 
 
When $T \gg 1$,  the path integral is dominated by the saddle-points of its total action, giving rise to effective Hamiltonian dynamics for the rare fluctuations. Therefore, in the $T \to \infty$ limit, the cumulant generating function $K(\gamma)$ is found by evaluating the average in \eqref{CGF} for $p(x,t)$ and  $q(x,t)$  that maximize the total action. We can thus write
\begin{equation}
   K(\gamma) =   \beta T^2 \mu(\lambda) = \beta T^2
   \mu\left(\frac{\gamma}{\beta T}\right) \, ,
   \label{Kmu}
\end{equation}
where the function $\mu(\lambda)$ is  given by the value of the functional evaluated at  $p(x,t)$ and $q(x,t)$  that maximize the total action
\begin{equation}
   \mu(\lambda) =  - S[q,p]  - {\mathcal F}[q]
   + \lambda  \int_0^\infty dx\,[q(x,1) - q(x,0)]. 
   \label{mu-optimal}
\end{equation}

The calculation of the function $\mu$ is one of the main objectives of this  work.

\subsection{Governing equations}

The optimization of the action leads to a pair of partial differential equations (PDEs) for the density $q(x,t)$ and momentum $p(x,t)$. These deterministic nonlinear PDEs follow from the single stochastic Kawasaki-Dean equation \eqref{DeanKawasaki}, but they are more manageable and constitute the basis of MFT, see \cite{Derrida07,MFT15} for review of the MFT in more simple diffusive lattice gases. The solution of the MFT equations gives the optimal path, viz. the density history providing a dominant contribution to the probability of observing a given current (or a rare fluctuation of other observable). The MFT is a powerful tool for probing large deviations in such systems, see, e.g., \cite{DG09,Meerson12a,Meerson14,Kirone14}. 

The MFT has been extended beyond the original class of diffusive lattice gases \cite{Bertini01,Bertini02,Derrida07,MFT15}. For the Dyson gas with long-range interactions, we must also extend the MFT. Indeed, only local terms usually appear, not just in diffusive lattice gases but also in diffusive systems with a few interacting scalar fields (see, e.g., \cite{Pablo19}), systems with conserved noise \cite{Vilenkin16}, etc. Non-local terms containing the Hilbert transform appear for the Dyson gas. The derivation of the governing equations is straightforward, we use the action \eqref{Action} and deduce the saddle-point equations (see Appendix~\ref{app:MFTpq} for details):
\begin{subequations}
\label{MFT:qp}
\begin{align}
  \partial_t q  &=  - \partial_x\left( 
  \pi q  \Hil[q] + 2 q \partial_x p  - \frac{C_\beta}{T}\, \partial_x q \right),
  \label{MFTq0} \\
  \partial_t p  &=  \pi  \Hil[q\partial_x p] - \pi   \Hil[q]\partial_x p
  - (\partial_x p)^2  - \frac{C_\beta}{T}\, \partial^2_{x} p \,.
   \label{MFTp0}
\end{align}
\end{subequations}

Equations \eqref{MFT:qp} describe the fluid with non-local interactions as they contain the non-local Hilbert transform. The MFT equations for the Riesz gases with a sufficiently small exponent, $0<s<1$, are also non-local \cite{RahulRiesz} as manifested by contain the generalized Hilbert transform defined by \eqref{def:Hilb} with $(x-y)^{-1}$ replaced by $(x-y)|x-y|^{-2-s}$. Non-local MFT equations also appear in Ginzburg-Landau dynamics with long-range interactions \cite{RaphCedric}.

Equations \eqref{MFT:qp} are obtained by optimizing the action \eqref{Action}. Therefore, the same equations arise in other rare events, not only the probability of observing an anomalously large current. The boundary conditions at the initial and final times reflect the nature of the specific rare event. We deduce these boundary conditions by optimizing other contributions to action.

The terms with $C_\beta$ in \eqref{MFT:qp} scale as $O(1/T)$, so they are subdominant when $T\gg 1$. Omitting these subdominant  terms we simplify Eqs.~\eqref{MFT:qp} to 
\begin{subequations}
\label{MFT:qp-simple}
\begin{align}
  \partial_t q  &=  - \partial_x\left(\pi q  \Hil[q] + 2 q \partial_x p\right),
  \label{MFTq} \\
  \partial_t p  &=  \pi  \Hil[q\partial_x p] - \pi   \Hil[q]\partial_x p
  - (\partial_x p)^2.
   \label{MFTp}
\end{align}
\end{subequations}

At the final time, $t=1$, the boundary condition reads 
\begin{equation}
p(x,1) = \lambda \theta(x)   \, . \label{BCfin}
\end{equation}
This condition has to be satisfied only in the region where $q(x,1) >0$ (see Appendix~\ref{app:MFTpq}). For $t=0$,  the boundary condition is 
\begin{equation}
  p(x,0) =     {\lambda} \theta(x) +
  \int_{-\infty}^{\infty} dy ~ (1 - q(y,0)) \ln{|x-y|}  \, .
  \label{InitAnnealed}
\end{equation}

The solution of  Eqs.~\eqref{MFT:qp-simple} subject to the boundary conditions \eqref{BCfin} and \eqref{InitAnnealed} represents the density profile history producing a given current. The CGF in the $T \to \infty$ limit is found by evaluating the formula \eqref{mu-optimal} for the optimal path. Because the total action is stationary along the optimal path, we can use the identity \cite{Vivo16,Meerson22a,Dandekar22} 
\begin{equation}
     \frac{ d \mu}{d \lambda} = \int_0^\infty dx \, [q(x,1) - q(x,0)]
     \label{Idmu}
\end{equation}
that considerably simplifies the computations.

\subsection{The complex Burgers equation}
\label{subsec:Burgers}

Here, we map Eqs.~\eqref{MFT:qp-simple} into the complex Burgers equation \cite{Matytsin, BlaizotQCD,C-Burgers,Blaizot,menonCOMPLEX} that is more suitable for further analysis. We begin by outlining a hydrodynamic interpretation that we also use for mapping into the complex Burgers equation. Following Matytsin \cite{Matytsin} (see also \cite{Instanton14}), we define a local velocity field 
\begin{equation}
  v  = \pi \Hil[q] + 2 \partial_x p \, , 
\end{equation}
and rewrite Eqs.~\eqref{MFT:qp-simple} as 
\begin{subequations}
\label{M12}
\begin{align}
	  \partial_t q + \partial_x \left( q v \right) &= 0\,,
          \label{eq:M1}\\
	\partial_t v + v (\partial_x v) &= \pi^2 q (\partial_x q)\,.
	\label{eq:M2}
\end{align}
\end{subequations}
Equation \eqref{eq:M1} plays the role of a continuity equation.  The right-hand side of \eqref{eq:M2} can be re-written in the hydrodynamic form, $-q^{-1}\partial_x P$ with $P=-\pi^2 q^3/3$, so Eq.~\eqref{eq:M2} resembles the Euler equation \cite{Landau-FM} describing a one-dimensional isoentropic flow of an ideal gas with adiabatic index $\gamma=3$. Recalling that $\gamma=1+2/d$ for the ideal monoatomic gas in $d$ dimensions \cite{Landau-FM,Beijeren}, the pressure $P = -\pi^2 q^3/3$ corresponds to the intrinsically one-dimensional monoatomic gas, although it has the `wrong' sign. The same pressure also appears in the context of one-dimensional free fermion gas \cite{Abanov06}. 

Pursuing the hydrodynamic analogy we recall that any one-dimensional flow is necessary a potential flow, so one writes $v=\partial_x \phi$, integrates \eqref{eq:M2} with respect to $x$, and arrives at a Bernoulli integral
\begin{equation}
\label{Bernoulli}
\partial_t \phi + \tfrac{1}{2}v^2+w=0
\end{equation}
The right-hand side in \eqref{Bernoulli} can be an arbitrary function of time. Without loss of generality, one can set it to zero since the potential is defined up to an arbitrary function of time as it does not affect the velocity $v=\partial_x \phi$. The enthalpy in \eqref{Bernoulli} is $w=-\pi^2 q^2/2$, i.e., it has again the `wrong' sign. Introducing an auxiliary function \cite{Landau-FM} 
\begin{equation}
\label{chi:def}
\chi = \phi - x v + t\left(\tfrac{1}{2}v^2+w\right)
\end{equation}
and treating it as a function of $v$ and $w$ one recasts \cite{Landau-FM} the continuity equation \eqref{eq:M1} into a linear PDE
\begin{equation}
\label{chi:eq}
\left(2w\partial_w^2 - \partial_v^2+ \partial_w\right)\chi(w,v) = 0
\end{equation}
Using the variable $u=\sqrt{2w}$ instead of $w$, one transforms \eqref{chi:eq} into the standard wave equation
\begin{equation}
\label{chi:wave}
\left(\partial_u^2 - \partial_v^2\right)\chi(u,v) = 0
\end{equation}
whose general solution is $\chi = \Omega_1(u+v)+\Omega_2(u-v)$ with arbitrary functions $\Omega_1$ and $\Omega_2$. 

In contrast to standard hydrodynamics where $u$ is real, the variable $u$ is now complex, more precisely, $u$ is purely imaginary: $u = \ii \pi q$. Hence, the usage of the complex variables seems inevitable. It is possible to push the above hydrodynamics-inspired approach, but instead, we perform the complexification from the beginning. We introduce the complex-valued function 
\begin{equation}
   f(x,t) = v +  \ii \pi q = \pi \Hil[q] +  2 \, \partial_x p  + \ii \pi q 
   \label{eq:Adef}
\end{equation}
and recast the pair of equations \eqref{M12} into a single complex Burgers equation
\begin{equation}
   \partial_t  f  + f \partial_x f =0 \, .
   \label{eq:ABurgers}
\end{equation}
A viscous $\nu\partial^2_x f$ term usually appearing in the Burgers equation is absent, so Eq.~\eqref{eq:ABurgers} is also known as an inviscid complex Burgers equation. The subdominant terms that we dropped from Eqs.~\eqref{MFT:qp-simple}, and hence they also do not appear in Eqs.~\eqref{M12}, provide the (asymptotically negligible) viscous contribution.

The complex Burgers equation arises in many subjects, see, e.g., \cite{Kenyon,Blaizot, Blaizot2, Burda14,Instanton14,ForresterGrela,menonLesser2012,menonCOMPLEX} and references therein. An (implicit) solution of the Burgers equation reads
\begin{equation}
   f(x,t) = f_1[x +(1-t)  f(x,t)] \, ,
   \label{SolBurgers1}
\end{equation}
where we shortly write $f_1(z)\equiv f(z,t=1)$. An alternative form of the solution, $f(x,t) = f_0[x -tf(x,t)]$ with $f_0(z)\equiv f(z,t=0)$, is traditionally used since the `boundary' condition $f_0(z)$ at the initial time is usually known. In the present case, however, boundary conditions are mixed, some at the initial time $t=0$ and others at the final time $t=1$, cf. \eqref{BCfin}.

To appreciate another difficulty in dealing with the complex Burgers let us look at a {\em simpler} linear PDE
\begin{equation}
\label{transport}
\partial_t  f  + c \partial_x f = 0
\end{equation}
If $c$ is real, Eq.~\eqref{transport} is a hyperbolic PDE, and the solution $f(x,t)=f_0(x-ct)$ is meaningful \cite{menonCOMPLEX}. If $c=\ii$, Eq.~\eqref{transport} is the Cauchy-Riemann equations in disguise, so it is an elliptic PDE; more generally, Eq.~\eqref{transport} is elliptic if $\text{Im}(c)\ne 0$. The complex Burgers equation \eqref{eq:ABurgers} is elliptic at point $(x,t)$ if $q(x,t)>0$ and hyperbolic if $q(x,t)=0$. However, the solution by the method of characteristics is still expected to work if $f(x,t)$ remains analytic in the neighbourhood of the real line at all times \cite{menonCOMPLEX}. The analysis in Sec.~\ref{sec:calculCGF} confirms the relevance of the regions where $q$ vanishes and where $q$ is positive. 

We digress and note that if the evolution of the Dyson gas is unconstrained, the momentum vanishes, $p = 0$, and the (average) density $\rho(x,t)$ satisfies the hydrodynamic equation 
\begin{align}
  \partial_t \rho  + \partial_x(\pi \rho  \Hil[\rho]) = 0.
  \label{rho-eq} 
\end{align}
Some solutions of Eq.~\eqref{rho-eq} can be read off the previous results by setting $p=0$. In Appendix~\ref{app:vacuum}, we derive an amusing scaling solution describing the expansion of the Dyson gas into a vacuum. The density profile of this solution is the diffusively expanding Wigner semi-circle.

\section{Large deviations of the current}
\label{sec:calculCGF}
  
 \subsection{Time-reversal symmetry}

For a system initially at equilibrium, the solution of the governing equations exhibits a time-reversal invariance. A similar property also occurs in the symmetric exclusion process \cite{Benichou22,MMS22},  the Kipnis-Marchioro-Pressutti model \cite{Meerson22a}, and in the weak-noise theory of the KPZ equation \cite{LeDoussalKraj1}. This symmetry becomes manifest by defining a new variable 
 \begin{equation}
   \bar{p}(x,t) = p(x,t) + \int_{-\infty}^{\infty} dy\,[q(y,t)-1] \log{|x-y|} \, .
   \quad \label{eq:defpbar1}
 \end{equation}
 Taking the derivatives one obtains 
 \begin{eqnarray}
   \partial_x \bar{p} -  \partial_x p  =  \pi \Hil[q-1]
   =  \pi \Hil[q]   \, . 
   \label{eq:defpbar2}
 \end{eqnarray}
Using the identity $ \Hil^2 = -1$, we invert \eqref{eq:defpbar2} and find
\begin{equation}
    q-1 = \frac{1}{\pi} \Hil[\partial_x p] -
    \frac{1}{\pi}\Hil[\partial_x \bar{p}]  \,. 
     \label{eq:qpbarp}
\end{equation}
The  boundary condition (\ref{InitAnnealed}) takes a very simple form in terms of $\bar{p}$: 
\begin{equation}
    \bar{p}(x,0) = \lambda \theta(x)    \, . \label{eq:boundnew}
\end{equation}
Re-writing Eqs.~\eqref{MFTq}--\eqref{MFTp} via $p$ and $\bar{p}$  display a symmetry relation (see Appendix \ref{app:PTsym} for the derivation)
\begin{equation}
 p(x,t) =    \lambda -  \bar{p}(-x,1-t)   \, .\label{eq:sympbarp}
  \end{equation}
Combining \eqref{eq:qpbarp} and \eqref{eq:sympbarp} we get
\begin{equation}
 q(x,t) = q(-x,1-t)\,.    
 \label{eq:symprofile}
 \end{equation}
Using \eqref{eq:defpbar2} and \eqref{eq:qpbarp} we find that the function $f$ defined by Eq.~\eqref{eq:Adef} can be written as 
\begin{equation}
  f(x,t) =  \partial_x p  +  \ii \Hil[\partial_x p]
  + \partial_x \bar{p}  -  \ii \Hil[\partial_x \bar{p}]  + \ii \pi \, .
  \label{eq:fppbar}
\end{equation}
From this expression we deduce that $f$ satisfies 
\begin{equation}
   f(x,t) =  f(-x,1-t) \, .
   \label{eq:symfPT}
\end{equation}
Equations \eqref{eq:symprofile} and \eqref{eq:symfPT} assert that the functions $q$ and $f$ obey the parity-time (PT) space-time reflection symmetry. (The PT symmetry is traditionally discussed in the context of quantum mechanics \cite{PT-Bender}, but it also arises in classical systems \cite{PT-El}.) Combining the PT symmetry with the implicit solution of the complex Burgers equation, Eq.~\eqref{SolBurgers1}, we can write
\begin{equation}
2  f(x,t) =  f_1[x + (1-t) f(x,t)] + f_1[-x + t f(x,t)].
   \label{SolBurgers2}
\end{equation} 

We now turn to another way of looking at Eq. \eqref{eq:fppbar} that allows us to separate the contributions of $p$ and $\bar{p}$. We express $f(x,t)$ as the sum of two functions
\begin{equation}
  f(x,t) = F_1(x,t) + F_2(x,t) 
  \label{fRH}
\end{equation}
with $F_1$ depending on $p$ and $F_2$ on $\bar{p}$:
\begin{subequations}
\begin{align}
  F_1(x,t) &=   \partial_x p  +  \ii \Hil[\partial_x p] + \frac{\ii \pi}{2} \, , 
  \label{def:F} \\
  F_2(x,t) &= \partial_x \bar{p}  -  \ii \Hil[\partial_x \bar{p}]
   + \frac{\ii \pi}{2}    \, . \label{def:G} 
\end{align}
\end{subequations}
These two functions have complementary properties in the complex plane. This property is important as the complex Burgers equation involves flows from the neighborhood of the real line.

A theorem due to Titchmarsh \cite{Titchmarsh48} asserts that a square-integrable complex-valued function on the real line, with an imaginary part given by Hilbert transform of its real part, can be extended to a holomorphic function in the upper complex half-plane. Hence, we can extend $F_1(x,t)$ to the function $F_1(z,t)$ analytic in the upper complex half-plane ($\text{Im}(z)>0$). Similarly,  we extend $F_2(x,t)$ to the function  $F_2(z,t)$ analytic in the lower complex half-plane. Equation \eqref{fRH} then represents a Riemann-Hilbert decomposition of the function $f$.  In terms of this decomposition, the boundary conditions \eqref{BCfin} and \eqref{eq:boundnew} lead to constraints
\begin{subequations}
\begin{align}
  \mbox{Re}[F_1(x,1)] &= \lambda \delta(x)  \, ,  \\
  \mbox{Re}[F_2(x,0)] &= \lambda \delta(x)  \, ,
\end{align}
\end{subequations}
 for $F_1(x,1)$ and $F_2(x,0)$. The PT symmetry \eqref{eq:symfPT} implies the general relation
\begin{equation}
  F_1(x,t) = F_2(-x,1-t) 
  \label{SymFG}
\end{equation}
valid for $x$ real. Equation \eqref{SymFG} allows us to focus on $F_1$. 

\subsection{Calculation of the density at the final time}

The function $F_1(z,t)$ is analytic in the upper half plane, and hence it can be represented in the resolvent form
\begin{equation}
  F_1(z,t) = \int_{-\infty}^{\infty} \frac{\rho_1(x,t)}{x -z} {\rm d}x \, .
\end{equation}
Using the Sokhotski-Plemelj relation (we denote by ${\mathcal P}$ the Cauchy principal value) 
\begin{equation}
  \lim_{\epsilon \to 0^+} \frac{1}{x - \ii \epsilon} =
      {\mathcal P}\left( \frac{1}{x}\right) + \ii \pi \delta(x) \, , 
      \label{Sokhotski-Plemelj}
  \end{equation}
and \eqref{def:F} we find that the density $\rho_1$ is given by
\begin{equation}
  \rho_1(x,t) = \frac{1}{\pi}\mbox{Im}[F_1(x,t)] = \frac{1}{\pi} \Hil[\partial_x p] + \frac{1}{2} \, .
\end{equation}
From this equation, we see that $\rho_1$ tends  to 1/2 as $x$ goes to infinity. Similarly, we define a density $\rho_2$ to represent $F_2(z,t)$ such that
\begin{equation}
  \rho_2(x,t) =  \frac{1}{\pi}\mbox{Im}[F_2(x,t)] =
  -\frac{1}{\pi} \Hil[\partial_x \bar{p}] + \frac{1}{2} \, . 
  \label{eq:rho2def}
\end{equation}
From Eq.~(\ref{eq:qpbarp}), we obtain the following decomposition of the
 density profile
\begin{eqnarray}
  q(x,t) &=& \rho_1(x,t) +  \rho_2(x,t) \nonumber \\
            &=&  \rho_1(x,t) +  \rho_1(-x,1-t) \, ,
 \label{decom-q}
\end{eqnarray}
where we have used the symmetry \eqref{SymFG} in the last equality. Using $\partial_x p = \mbox{Re}[F_1] = -\pi \Hil[\rho_1]$,
we recast the boundary condition \eqref{BCfin} into equation 
\begin{equation}
  \int_{-\infty}^x dy\,  \Hil[\rho_1](y,1) = - \frac{\lambda}{\pi}\theta(x) 
  \label{eq:rhof}
\end{equation}
applicable in the region where $\rho_1(x,1)>0$. Equation \eqref{eq:rhof} can be interpreted as the problem of finding the equilibrium density profile of a Dyson gas subject to a step potential. Differentiating \eqref{eq:rhof}  with respect to $x$ gives
\begin{equation}
  \Hil[\rho_1](x,1) = - \frac{\lambda}{\pi}\delta(x)
  \quad {\rm when } \quad  \rho_1(x,1) > 0  \, .
  \label{eq:rhof2}
\end{equation}
Assuming that $\rho_1 >0 $ everywhere leads to a contradiction. Indeed, inverting the Hilbert transform \eqref{eq:rhof2}, we obtain $\rho_1(x,1) =  \frac{1}{2} + \frac{\lambda}{\pi^2 x}$ that becomes negative for $x \to 0^-$ and is not integrable when $x \to 0^+$,  so it cannot be a density. Hence, Eq.~\eqref{eq:rhof} does not hold near 0: this can only happen if  $\rho_1(x,1)$   vanishes in the vicinity of the origin. 

A similar situation occurred in \cite{majumdarHow2011} devoted to the stationary density of the Dyson gas with harmonic confining potential supplemented by the step potential. In this setting, the density vanishes on an interval and exhibits a square-root divergence near the upper bound of that interval. In our situation, we expect the square-root divergence to occur on the right of the origin since we consider a large positive current. These considerations suggest to seek a solution to Eq.~\eqref{eq:rhof} satisfying 
\begin{enumerate}
	\item $\rho_1(x,1) \rightarrow \frac{1}{2}$ as $|x| \rightarrow \infty$
	\item  $\rho_1(x,1) = 0$ in a region $(-a,0)$. 
	\item $\rho_1(x,1)$ has an integrable singularity
          as  $x \rightarrow 0^{+}$.
	\item  Using  \eqref{eq:rhof} for $x$  between $-a$ and $0^{+}$  implies
	  \begin{equation}
	    \int_{-a}^{0^+} dx\,   \Hil[\rho_1](x,1) =  - \frac{\lambda}{\pi} \, .
            \label{jumpcond}
	\end{equation}
\end{enumerate}
Guided by the confined case \cite{majumdarHow2011}, we seek a solution of the form $\rho_1(x,1) = \frac{1}{2} \sqrt{1+\frac{a}{x}}$ for $x>0$ and $x<-a$, that vanishes when  $ x \to -a^{-}$ and has a square-root divergence for $x \to  0^{+}$. The Hilbert transform of this density $ - \frac{1}{2}  \sqrt{1+\frac{a}{y}}$ for $-a<y<0$. Imposing the condition \eqref{jumpcond} we obtain $a = 4 \lambda/\pi^2$, and therefore 
\begin{eqnarray}
\rho_1(x,1) = 
\begin{cases}
     \frac{1}{2} \sqrt{1 + \frac{4 \lambda}{\pi^2 x}}  & \mbox{ for } x \notin \left[- \frac{4\lambda}{\pi^2},0\right]\\
     0 &\mbox{ for }  x \in \left[- \frac{4\lambda}{\pi^2},0\right]
\end{cases}
  \label{eqsol:rho1fin}
 \end{eqnarray}
Using again the Sokhotski-Plemelj formula  \eqref{Sokhotski-Plemelj}, equation~\eqref{eqsol:rho1fin}
 leads to 
\begin{equation}
  F_1(z,1) = F_2(-z,0) = \ii \frac{\pi}{2} \sqrt{1 + \frac{4 \lambda}{\pi^2 z}} \, . 
  \label{eq:Fsolann}
 \end{equation}

Finally, we determine  $\rho_2(x,0)$.  From Eq.~\eqref{SolBurgers2} at $t=1$, we obtain
an implicit equation for $f_1$:
\begin{equation}
    f_1(x) = f_1(-x+ f_1(x))  \, .  \label{eq:fcond1}
\end{equation}
We know from Eq.~\eqref{fRH} that
\begin{equation}
	f_1(x) = F_1(x,1) + F_2(x,1)  \, , \label{eq:fcond2}
\end{equation}
where $F_1(x,1)$ is given by \eqref{eq:Fsolann}. The form of $F_2(x,1)$ is implicitly specified by the requirement that it satisfy \eqref{eq:fcond1}, and the fact that $F_2$ is analytic in the lower half plane. Now, since $f_1(x)$ must have the `symmetry' \eqref{eq:fcond1}, we write a suitably `symmetrized' ansatz for $f_1$:
\begin{equation}
  f_1(x) = \ii \frac{\pi}{2} \sqrt{1+\frac{4 \lambda}{\pi^2 x}} + \ii \frac{\pi}{2} \sqrt{1 + \frac{4 \lambda}{\pi^2 (-x + f_1(x))}}
  \label{Ansatzf1}
\end{equation}
We can use \eqref{Ansatzf1} provided that the second term is analytic in the lower half plane, allowing us to identify it with $F_2(x,1)$. However, since this is still an implicit equation for $f_1$, it is difficult to handle. We thus adopt a perturbative approach in $\lambda$ for analyzing \eqref{Ansatzf1}.

At the lowest order, we have $p^{(0)}=0$ and $q^{(0)}=1$, leading to $f_1^{(0)} = \ii \pi$. This zero-order result follows from \eqref{Ansatzf1} by taking $\lambda \to 0$  for fixed values of $x$. At the next order
\begin{equation}
    f_1^{(1)}(x) = \ii \frac{\pi}{2} \sqrt{1+\frac{4 \lambda}{\pi^2 x}} + \ii \frac{\pi}{2} \sqrt{1 +  \frac{4 \lambda}{\pi^2 (-x + \ii\pi)}} \, .
\end{equation}
Observing that the second term is  analytic in the lower half plane, we obtain  $F_2(x,1)$ at the  dominant order in $\lambda$.
Using Eq.~\eqref{eq:rho2def}, only keeping terms that are relevant to $K(\lambda)$ in the $T\to\infty$ limit (see \eqref{Kmu}), 
we get 
\begin{equation}
   \rho_2(x,1) = \frac{1}{2} - \frac{\lambda}{\pi^2} \frac{x}{x^2 + \pi^2} \, .
\label{eqsol:rho0init}
\end{equation}

\subsection{The cumulant generating function (CGF)} 

From the expression \eqref{Idmu} for the derivative of the CGF and the decomposition \eqref{decom-q},  we deduce 
\begin{equation}
   \mu'(\lambda) = 2 \int_0^\infty dx \left[\rho_1^\text{odd}(x,1)+ \rho_2^\text{odd}(x,1)\right]
 \label{Idmu2} 
\end{equation}
where $\rho_\ell^\text{odd}$ with $\ell=1, 2$ is the odd part:
\begin{equation*}
2\rho_\ell^\text{odd}(x,1)=\rho_\ell(x,1) - \rho_\ell(-x,1). 
\end{equation*}

Substituting $\rho_1$ and $\rho_2$ at the final time, \eqref{eqsol:rho1fin} and \eqref{eqsol:rho0init}, into Eq.~\eqref{Idmu2} and evaluating the integrals we derive the CGF of the total current  
\begin{equation}
  \mu(\lambda) = \frac{\lambda^2}{\pi^2}\left(
  \frac{1}{2}\log\frac{\pi^6}{\lambda^2} + \frac{1}{2} \right)  \, . 
  \label{formuleCGF}
\end{equation}
We have written $\mu(\lambda)$ as a manifestly even function of $\lambda$. 

Using Eqs.~\eqref{Legendre1}, \eqref{Kmu} and \eqref{formuleCGF} we obtain the distribution of the current 
\begin{eqnarray}
  \log [{\rm Prob}(J)] &=& \beta T^2
  \max_\lambda \left( \mu(\lambda)  - \lambda \frac{J}{T} \right)
   \nonumber \\
  &=& \beta T^2\left( \mu(\lambda^*)  - \lambda^* \frac{J}{T} \right)
 \label{distributionJ}
\end{eqnarray}
where $\lambda^*$ satisfies
\begin{equation}
  \mu'(\lambda^*) =  \frac{2 \lambda^*}{\pi^2}
  \log\frac{\pi^3}{\lambda^*} =  \frac{J}{T} \, .
  \end{equation}
This equation is solved in terms of the  Lambert function 
\begin{equation}
  \lambda^* =  - \frac{ \pi^3 j}{W_{-1}(-j) } \,, \quad j \equiv \frac{ J}{2 \pi T}\,.
\end{equation}
We have chosen the real branch $W_{-1}$  of the Lambert function, as this ensures that $\lambda \to 0$ when $J \to 0$. Substituting this value of $\lambda^*$ in \eqref{distributionJ}, we obtain the probability distribution of the current 
\begin{equation}
 \log[{\rm Prob}(J)] = \frac{\pi^2 \beta J^2/4}{W_{-1}(-j)} \left[1+\frac{1}{2W_{-1}(-j)} \right].
  \label{pdf-exp}
\end{equation}
In the $T \to \infty$ limit, we have $j\to 0$. The asymptotic $W_{-1}(-j) \simeq \log j$ of the Lambert function leads to
\begin{equation*}
W_{-1}(-j) \simeq  \log \left(\frac{ J}{2 \pi T} \right)  \simeq 
 - \tau, \quad \tau\equiv \log 2\pi T
\end{equation*}
and allows us to approximate the probability density function \eqref{pdf-exp} by Gaussian
\begin{equation}
  {\rm Prob}(J) \simeq
 \sqrt{\frac{\pi \beta}{4 \tau}}\,
  \exp\!\left(- \frac{\pi^2 \beta J^2}{4 \tau  } \right),
\end{equation}
from which the variance of the current is
\begin{equation}
  \langle J^2 \rangle  = \frac{2}{ \pi^2 \beta } \log(2\pi T) \simeq 
  \frac{2 D}{\pi^2 g}\log( g \rho_0^2 t_f)  
  \label{Variance-Log}
\end{equation}
in the $T\to\infty$ limit. Fluctuations of the current are drastically reduced compared to single-file diffusion where the variance scales as $t^{1/2}$ \cite{Diamant, DG09,Barkai13,Benichou13, Kirone15a, Benichou22}. Long-range logarithmic interactions between the particles thus almost `freeze' the motion of the particles: the Dyson fluid is effectively incompressible. Fluctuations of the current are also reduced compared to the single-file fluctuations for the Riesz gas with index $0<s<1$. The variance of the current scales as  $t^{s/(s+1)}$, see \cite{RahulRiesz}, suggesting the logarithmic behavior since the Dyson gas corresponds to the  $s \to 0$ limit.  

The growth law \eqref{Variance-Log} agrees with the expression obtained by Spohn \cite{Spohn87,Spohn87b, Spohn98MPRF} via a mapping to free fermions that is valid for $\beta =2$. The Coulomb gas method \cite{dean06} and the hydrodynamic approach allowed us to consider arbitrary values of $\beta$ and to compute the higher-order cumulants of the integrated current.

\subsection{Higher cumulants} 
 
We should go beyond the Gaussian approximation for the probability density function to calculate higher-order cumulants. Using the asymptotic
\begin{equation*}
W_{-1}(-j) \simeq -\log(1/j) - \log[\log(1/j)]
\end{equation*}
valid when $0<j\ll 1$ we approximate the probability distribution \eqref{pdf-exp} by
\begin{equation}
  \log[{\rm Prob}(J)] \simeq -\frac{\pi^2 \beta J^2}{4\tau}\left[1-\frac{\log(\tau/J)+\frac{1}{2}}{\tau}\right]
   \label{pdf-Expr2}
\end{equation}
which we use to calculate the moments and the cumulants of the current in the large time limit (see Appendix~\ref{app:cumulants} for details of the calculations). For $m \ge 2$, the cumulant of order $2m$ is 
\begin{eqnarray}
   \langle J^{2m} \rangle_c \simeq \frac{(m-2)!}{\pi^2 \beta}
   \left(-\frac{ 4 \log  T}{ \pi^2 \beta} \right)^{m-1}  
   \label{eq:formulecumulant}
\end{eqnarray}
in the leading order. The ratio $\langle J^{2m} \rangle_c/\langle J^{2} \rangle^m$ vanishes in the long time limit. Therefore the fluctuations of the current are Gaussian in the leading order. The same feature is true for the statistics of the number of eigenvalues in a large interval of an orthogonal, unitary, or symplectic random matrix \cite{CostinLebowitz}.

\subsection{Dyson fluid and other subjects}

Equations \eqref{M12} describing Dyson `fluid' appear, under different guises, in numerous disparate subjects. We have already mentioned the mapping into the complex Burgers equation \cite{Kenyon,Blaizot, Blaizot2, Burda14,Instanton14,ForresterGrela,menonLesser2012,menonCOMPLEX} arising, e.g., in the context of random matrices, limit shapes, and quantum chromodynamics. The complex Burgers equation has connections with the Calogero-Moser system (particles on the line interacting via the $r^{-2}$ potential \cite{Calogero69,Alexios06,Dhar19,Touzo24}), the Harish-Chandra-Itzykson-Zuber integral \cite{Matytsin,Instanton14,menonCOMPLEX,Bouchaud}, etc. 

The similarity of the Dyson fluid with other subjects is also reflected in similar phenomenology. For instance, the logarithmic growth law \eqref{Variance-Log} appears in 1D free fermion systems \cite{Tibor08,Moriya_2019} and in the spectral number fluctuations in random matrix theory \cite{CostinLebowitz,foglerProbability1995,DysonBOOK,Mehta,Forrester,majumdarHow2011,marinoNumber2016}. 

We also mention a striking connection between the Dyson gas and stochastic particle systems with localized interactions \cite{Popkov1,Popkov2}. For instance, the asymmetric simple exclusion process conditioned to carry an atypically large current \cite{Simon2009,RaphHugo,RaphHugo2, meerson-extreme} becomes asymptotically equivalent to the free fermion XX model in imaginary time. In other words, imposing an atypically current generates long-range interactions, and the system becomes identical to the Dyson gas. More precisely, the phase diagram for atypical behavior has two regions \cite{Karevski}. The conformally invariant region arises when the current is atypically large (the Dyson gas emerges when the current diverges). A phase separation region corresponds to atypically low current.  Along the critical line, the Kardar-Parisi-Zhang universality class holds \cite{Karevski}. This example shows that non-equilibrium processes under extreme conditions may become equivalent to the Dyson gas (see also  \cite{ZahraAli}).

\section{Conclusions}
\label{sec:concl}

We investigated the stochastic Dyson gas: $N$ point particles performing overdamped motion caused by repulsive interactions through a logarithmic potential with strength $g$ and independent identical Gaussian white noises acting on particles. When $N=1$, the particle undergoes standard Brownian motion with diffusion coefficient $D$. The dimensionless parameter $\beta= g/D$, the characteristic ratio of deterministic to stochastic contributions, can be an arbitrary positive number. For three exceptional values, $\beta=1,2,4$, the system is identical to the Dyson Brownian motion performed by the eigenvalues of a (symmetric, hermitian, or symplectic, respectively) random matrix whose matrix elements are independent Brownian processes. 

In the large $N$  limit, the Dyson gas admits a continuous description: The Dyson fluid satisfies fluctuating hydrodynamics. Long-range interactions strongly constrain the motion of individual particles and make the fluid very rigid (incompressible). Quantitatively, the rigidity is reflected by fluctuations exhibiting anomalously slow logarithmic growth. 

The techniques developed in non-equilibrium statistical mechanics to investigate large deviations in diffusive gases \cite{Spohn91,Bertini01,MFT15} can be adapted to the Dyson gas. Again, one must again solve a pair of nonlinear coupled PDEs which are now non-local in contrast to the PDEs underlying the MFT of diffusive lattice gases  \cite{Bertini01, MFT15}. We derived these equations in the specific problem conditioned on atypically large current. The same equations describe the emergence of other rare events in the Dyson gas. The specificity of the problem is reflected in the conditions at the initial time and the final time. 

In the case of the current, the annealed (or random) setting when the Dyson gas is prepared at equilibrium is more tractable than the quenched (or deterministic) setting when the initial condition is fixed. The technical reason for the tractability of the annealed case is that the problem enjoys the parity-time (PT) space-time reflection symmetry that we used to determine, at leading order in the large time limit,  the probability distribution of the current and all its cumulants. If the initial density is fixed at $t =0$ and not allowed to fluctuate (quenched setting), the governing equations are the same as in the annealed case and the final condition at $t=1$ is still given by (\ref{BCfin}). However, we must replace the  initial time condition  \eqref{InitAnnealed} by 
 \begin{equation}
  q(x,0) = 1 \, .
  \label{InitQuench}
\end{equation}
This seemingly innocent modification spoils the PT invariance of the boundary-value problem, rendering the analysis much more challenging. In the quenched case, we only calculated the asymptotic behavior of the variance of the current 
\begin{equation}
  \langle J^2 \rangle_\text{quenched} = \frac{1}{\pi^2 \beta} \log{T}
   = \frac{1}{2} \langle J^2 \rangle_\text{annealed}
\end{equation}
Therefore the variance in the quenched setting is twice smaller than the annealed case. 

The difference in large deviations in the annealed and quenched settings arises in various lattices gases, particularly in one dimension \cite{DG09, Derrida09BA, Meerson12a, Barkai13,Meerson14} where the discrepancy tends to be everlasting. In the one-dimensional Riesz gas \cite{RahulRiesz}, the ratio of the variances is $2^{-\frac{1}{s+1}}$ when $0<s<1$. When $s>1$, the interaction is effectively short range, and the ratio stabilizes to $1/\sqrt{2}$. The determination of the quenched CGF for the Dyson gas remains, at the moment, a challenging open problem.

The statistical properties of a tagged particle can be readily related to current fluctuations when particles cannot cross and remain in the same order \cite{Varadhan2013,Imamura}. This property is often built in stochastic lattice gases such as exclusion processes. In the case of the Dyson gas, particles do not cross with probability one when $\beta \ge 1$ (see Appendix~\ref{app:crossing}). Let $X(t)$ be the displacement of the tagged particle at time $t$. The variance $\langle J^2 \rangle$ of the current is $\rho_0^2$ times larger than the variance $\langle X^2\rangle$ of the displacement as the typical distance between particles is $\rho_0^{-1}$. Thus in the annealed case
\begin{equation}
  \langle X_t^2 \rangle
   = \frac{1}{\rho_0^2} \langle J^2 \rangle = \frac{1}{\pi^2 \rho_0^2 \beta} \log{T}
\end{equation}
and we recover the result of \cite{Spohn87}. For $\beta < 1$, the ordering is not conserved by the dynamics. We believe that the tagged particle should exhibit normal diffusion; the precise value of the self-diffusion coefficient is unknown.

We emphasize again striking similarities between the dynamics of classical Dyson gas and the quantum dynamics of one-dimensional fermions
\cite{SchutzGM,Abanov06,Tibor08,Stephan_2011,Stephan_2017,Krapivsky_2018,Moriya_2019,Caux20,Yeh_2022,pallister2024}. Qualitative behaviors observed in the classical domain usually arise in the quantum domain and vice versa. The logarithmic growth of the variance of the current occurs both for the Dyson gas and for the chain of free spinless lattice fermions, but slowly decaying oscillations were additionally observed \cite{Tibor08} in the latter case and not yet identified for the Dyson gas (although they may emerge in the quenched setting). In the Dyson gas, the equilibrium two-point correlation function exhibits the oscillatory $x^{-4/\beta} \cos(2\pi x)$ decay \cite{Forrester93,Forrester} when the `temperature' is sufficiently low, $\beta\geq 2$. The temporal oscillations in the Dyson gas may also arise only when $\beta\geq 2$.

There are strong arguments \cite{Forrester93,Forrester,Lelotte} in favor of  the appearance of a Berezinskii-Kosterlitz-Thouless (BKT) phase transition in the Dyson gas at equilibrium at $\beta=2$. Besides, we observed that in the fluctuating hydrodynamic equation~(\ref{DeanKawasaki}), the coefficient  $C_\beta$ changes its sign at $\beta =2$ (see Appendix~\ref{app:DKequation}). Exploring the dynamical signatures of the BKT phase transition is thus an intriguing challenge for future research.

Finally we emphasize that although equilibrium characteristics of Riesz gases in $d\geq 2$ dimensions have been rigorously investigated in the mathematical literature, see  \cite{Lewin22,Serfaty17,Serfaty18,Petrache20} and references therein, much less in known about the dynamical behavior.  The generalization of the fluctuating hydrodynamic approach to higher dimensions is an open field.

\bigskip
\noindent{\bf Acknowledgments.}
KM thanks H. Spohn for suggesting to work on the Dyson gas and S. Mallick for careful reading of the manuscript. KM also benefitted from discussions with 
N. Demni, G. Sch\"utz, and A. Zahra. The work of KM is supported by the project RETENU ANR-20-CE40-0005-01 of the French National Research Agency (ANR). RD and PLK are grateful to IPhT for excellent working conditions.

\appendix

\section{The Dean-Kawasaki equation}
\label{app:DKequation}

Here we derive the Dean-Kawasaki equation \eqref{DeanKawasaki} for the empirical density $q(x,t)$ defined in \eqref{defq}. We  follow the original paper \cite{dean96} and the recent exposition \cite{Touzo23}, see also \cite{Archer2004,frusawaControversy2000} for more mathematically inclined  discussions.  We take  $f$ to be a smooth test-function and define  
\begin{equation*}
F(\vec{x}(t)) := \frac{1}{T}  \sum_{i=1}^N f(x_i(t))  = \int {\rm d}x f(x) q(x,t)  \, .
\end{equation*}
Using Ito's formula we compute the derivative
\begin{equation*}
\frac{ d F(\vec{x}(t)) }{d t } =  \frac{1}{T}  \sum_{i=1}^N f'(x_i(t)) \dot{x_i}(t) + \frac{1}{ \beta T^2}  \sum_{i=1}^N f''(x_i(t))  \, . 
\end{equation*}
This equation is equivalent to   
\begin{eqnarray}
  \int dx\, f(x) \partial_t q(x,t) &=&
   \frac{1}{T^2}  \sum_{i=1}^N f'(x_i(t)) \sum_{j \neq i} \frac{1}{x_i-x_j}  
   \nonumber \\
   &+&  \frac{1}{ \beta T} \int {\rm d}x f(x) \partial_{xx} q(x,t)
    \nonumber \\
    &+&  \frac{1}{T} \sum_{i=1}^N f'(x_i(t))  \frac{ \eta_i(t)}  
    {\sqrt{\beta T}} \, .
    \label{A-Ito}
\end{eqnarray}
The first term on the right-hand side of \eqref{A-Ito} is transformed as \cite{RogersShi}
\begin{eqnarray}
  &&\sum_{i=1}^N f'(x_i(t)) \sum_{j \neq i} \frac{1}{x_i-x_j}  
   =\sum_{i=1}^N \sum_{j \neq i} \frac{f'(x_i(t)) - f'(x_j(t))}{2(x_i-x_j)} 
   \nonumber \\
   &&=   \frac{1}{2}  \sum_{i,j=1}^N  \frac{f'(x_i(t)) - f'(x_j(t))}{x_i-x_j}
   -   \frac{1}{2}    \sum_{i=1}^N  f''(x_i(t))   \nonumber \\
   &&= \int dx\,dy\, \frac{f'(x) - f'(y)}{2(x - y)}q(x,t)q(y,t)
   - \int dx \frac{f''(x) q(x,t)}{2}    \nonumber \\
   &&= \int {\rm d}x f'(x) q(x,t)     \dashint {\rm d}y \frac{q(y,t)}{x - y} 
   -  \int dx\, \frac{f(x) \partial_{x}^2 q(x,t)}{2}  \nonumber \\
   &&= -  \int {\rm d}x f(x)
   \partial_x   \left\{q(x,t) \pi  \Hil[q](x,t) + \frac{\partial_{x} q(x,t)}{2}
     \right\} 
   \label{Transfo1}
\end{eqnarray}
The last term on the right-hand side of \eqref{A-Ito} multiplied by $\sqrt{\beta T^3}$ can be written as  
\begin{eqnarray}
\sum_{i=1}^N f'(x_i(t))   \eta_i(t)
   &=& \int dx\,f'(x)
  \sum_{i=1}^N  \delta(x_i(t) - x)\eta_i(t)  \nonumber \\
  &=& - \sqrt{\beta T} \int dx\, f(x) \partial_x \Xi(x,t) 
  \label{Transf2}
\end{eqnarray}
with
\begin{eqnarray}
  \Xi(x,t) := \frac{1}{\sqrt{\beta T}}
   \sum_{i=1}^N  \delta(x_i(t) - x) \eta_i(t) \, . 
\end{eqnarray}
The last step is the calculation of the covariance of the stochastic noise $\Xi(x,t)$:
\begin{eqnarray}
   &&  \langle    \Xi(x,t)   \Xi(x',t') \rangle \nonumber  \\
   && = \frac{1}{\beta T} \sum_i\sum_j  \langle   \delta(x_i(t) - x) \eta_i(t)
   \delta(x_j(t') - x') \eta_j(t)\rangle  \nonumber  \\
   && = \frac{2}{\beta T}  \sum_i\sum_j  \delta_{ij} \delta(t - t')
   \delta(x_i(t) - x)  \delta(x_j(t') - x')  \nonumber  \\
   && = 2  \frac{q(x,t)}{\beta} \delta(t - t') \delta(x - x') 
   \label{CovXi}
\end{eqnarray}
 Hence,  the  noise  $\Xi(x,t)$ can be expressed as
\begin{equation}
  \Xi(x,t) = \sqrt{ \frac{q}{\beta}} \eta(x,t) \, , 
  \label{Xivseta}
  \end{equation}
where $\eta(x,t)$ is the Gaussian white noise satisfying \eqref{STWhite}. Substituting \eqref{Transfo1}, \eqref{Transf2} and \eqref{Xivseta} into \eqref{A-Ito} we arrive at the Dean-Kawasaki equation \eqref{DeanKawasaki} and also fix the amplitude $C_\beta = \frac{2 - \beta}{2\beta}$.  The change of sign of $C_\beta$ signals that something happens at $\beta=2$, and there are indeed arguments \cite{Forrester93,Forrester,Lelotte} that the BKT phase transition occurs in the Dyson gas at $\beta=2$.

\section{Derivation of the action and the governing equations}
\label{app:MFTpq}

Here we sketch the derivation of Eqs.~\eqref{MFTq0}--\eqref{MFTp0}. The path integral technique \cite{Cyrano,Janssen,MSR}, also known as Martin-Siggia-Rose formalism, allows us to write the exponential average of the current as 
\begin{subequations}
\begin{align}
\label{J:exp}
\langle {\rm e}^{\gamma J} \rangle &= \int {\mathcal D}q
         {\mathcal D}q_0  {\mathcal D}\eta\,\exp[\mathcal{J}]\, \delta(\partial_t q - \partial_x K)\\
\label{J:def}
\mathcal{J} & =    \lambda J - T^2 \beta {\mathcal F}[q_0]-\frac{1}{4} \int_0^1 dt\int_{-\infty}^\infty dx\,  \eta^2\\
\label{K:def}
  K                & = \frac{C_\beta}{T}\, \partial_x q -\pi q  \Hil[q ] +\frac{1}{T}\sqrt{\frac{q  }{\beta}} \,\, \eta(x,t)
  \end{align}
\end{subequations}
The term with delta-function in \eqref{J:exp} ensures the validity of the Dean-Kawasaki equation \eqref{DeanKawasaki}. The initial value of the density $q(x,0) = q_0(x)$ is randomly chosen with respect to the equilibrium measure (hereafter,  ${\mathcal D}q_0$ will be absorbed into  ${\mathcal D}q$). Rewriting the $\delta$ function with the help of  a Lagrange multiplier $p(x,t)$, we obtain
\begin{eqnarray}
    \langle {\rm e}^{\gamma J} \rangle =   \int {\mathcal D}q  {\mathcal D}p
        {\mathcal D}\eta\, \exp\!\big[\mathcal{J} - \int p(\partial_t q - \partial_x K)\big] 
\end{eqnarray}
with $\mathcal{J}$ defined by Eq.~\eqref{J:def}. The integration by parts with respect to the (unbounded) space variable $x$ gives $- \int dx\,p \partial_x K =\int dx\,K  \partial_x p $. Performing the Gaussian integral with respect to the noise $\eta$ yields
\begin{eqnarray}
  &&   \langle {\rm e}^{\gamma J} \rangle =
  \int {\mathcal D}q  {\mathcal D} p\,
       e^{T^2 \beta \left( \lambda \int dx\,[q(x,1) - q(x,0)] -  {\mathcal F}[q_0]
         - S[p,q]   
    \right) }\nonumber   
\end{eqnarray}
with bulk action $S[p,q]$ given by \eqref{Action}. In the $T \to \infty$ limit, the path integral is dominated by its saddle point, and at the level of exponentially dominant terms, we retrieve the expression \eqref{mu-optimal} for the CGF $\mu(\lambda)$.

Equations \eqref{MFTq0}--\eqref{MFTp0} are obtained by performing the variations $p \to p + \delta p$ and $q \to q + \delta q$ of the bulk action $S[p,q]$.  The boundary conditions  at the initial and final time  \eqref{InitAnnealed} and \eqref{BCfin} are derived by performing the variations at those moments: $q(x,0) \to q(x,0) + \delta q(x,0)$  and $q(x,1) \to q(x,1) + \delta q(x,1)$. [One performs integration by parts with respect to the (bounded)  time variable: $\int_0^1 dt\, p \partial_t q = p(x,1)q(x,1) - p(x,1)q(x,1) - \int_0^1 dt\, q \partial_t p$].

Because the density is positive, we can perform variation only in the region where $q>0$. Although this situation does not usually occur in the study of diffusive lattice gases \cite{MFT15}, it is well known that in random matrix theory (and thus for the Dyson gas) where solutions for density distributions can involve disjoint intervals. Therefore, the boundary conditions that we have derived should be obeyed only in regions where $q > 0$.

\section{Unconstrained evolution of the Dyson gas}
\label{app:vacuum}

Here, we step away from our chief goal, analyzing the evolution of the Dyson gas constrained on anomalously large current, and discuss the unconstrained evolution governed by the hydrodynamic equation \eqref{rho-eq} for the average density $\rho(x,t)$. We can formally solve Eq.~\eqref{rho-eq} by specializing the results of Sec.~\ref{subsec:Burgers} to $p = 0$ and writing $\rho$ instead of $q$. The function $f$ defined by \eqref{eq:Adef} becomes
\begin{equation}
   f(x,t) = \pi \Hil[\rho] + \ii \pi \rho
   \label{eq:f-rho}
\end{equation}
It satisfies again the complex Burgers equation \eqref{eq:ABurgers} and an ordinary initial condition $f_0(x)=\pi \Hil[\rho_0] + \ii \pi \rho_0$ with $\rho_0(x)=\rho(x, t=0)$. The solution is implicitly given by  
\begin{equation}
   f(x,t) = f_0[x -t f(x,t)] \, . 
   \label{Sol-B}
\end{equation}
The most profound feature of the (inviscid real) Burgers equation is the emergence of the shock waves \cite{Whitham}. The appearance of the shock waves in the realm of the complex Burgers equation is a more subtle issue discussed, e.g., in Refs.~\cite{BlaizotQCD,C-Burgers}. 

In this Appendix, we derive the fundamental solution describing the expansion into the vacuum. One can rely on the complex Burgers equation \cite{Blaizot}. We employ a more straightforward procedure that can be easily adapted to describe the expansion into the vacuum in the system with harmonic confining potential. 

The governing hydrodynamic equation (for the free system without external potential) reads 
\begin{equation}
\label{HD-Dyson}
\partial_t \rho(x,t) + \partial_x\!\left[g \rho(x,t)\dashint_{-\infty}^\infty dy\,\frac{\rho(y,t)}{x-y}\right]= 0
\end{equation}
in the original (dimensionful) variables. The fundamental solution describes the expansion of the Dyson gas into vacuum that begins with particles concentrated in a tiny spatial region. We thus postulate that all particles are initially at the origin and use 
\begin{equation}
\label{IC}
\rho(x,0)=N\delta(x)
\end{equation}
as the initial condition. The total number of particles is conserved throughout the evolution implying that
\begin{equation}
\label{mass}
\int_{-\infty}^\infty dx\,\rho(x,t)=N
\end{equation}
at all $t\geq 0$. 

The initial-value problem \eqref{HD-Dyson}--\eqref{IC} is invariant under the one-parameter group of transformations
\begin{equation}
\label{group}
\rho\to a^{-1} \rho, \qquad x\to a x, \qquad t\to a^2 t.
\end{equation}
This implies that $\sqrt{t} \rho(x,t)$ depends on the single variable $x/\sqrt{t}$. A carefully chosen self-similar form 
\begin{equation}
\label{scaling}
\rho(x,t) = \frac{N}{\sqrt{Ng t}}\,F(X)\quad \text{with} \quad X = \frac{x}{\sqrt{N g t}}
\end{equation}
ensures that both $N$ and $g$ disappear from the governing equation and the conservation law for the scaled density $F(X)$. The conservation law \eqref{mass} becomes
\begin{equation}
\label{F:IC}
\int_{-\infty}^\infty dX\, F(X) = 1.
\end{equation}
Substituting the scaling ansatz \eqref{scaling} into \eqref{HD-Dyson} we find that the scaled density satisfies
\begin{equation}
\label{F-eq-long}
F + X\,\frac{dF}{dX} = 2 \frac{d}{dX} \left[F(X) \dashint_{-\infty}^\infty dY\,\frac{F(Y)}{X-Y}\right]
\end{equation}
We now integrate \eqref{F-eq-long} and take into account the symmetry, $F(X)=F(-X)$, and vanishing of the scaled density in the $|X|\to\infty$ limits. The solution has compact support, $F=0$ for $|X|>R$; when $|X|<R$, the scaled density satisfies 
\begin{equation}
\label{F-eq}
X = 2\dashint_{-R}^R dY\,\frac{F(Y)}{X-Y}\,.
\end{equation}
The solution of the integral equation \eqref{F-eq} appears in textbooks and articles, e.g., \cite{Estrada,Carrillo17}. Summarizing
\begin{equation}
\label{F-sol}
F = 
\begin{cases}
\frac{1}{2\pi}  \sqrt{R^2-X^2}  & |X|<R\\
0                                       & |X|>R
\end{cases}
\end{equation}
The conservation law \eqref{F:IC} fixes $R=2$. In the original variables, the density profile is the Wigner semi-circle 
\begin{equation}
\label{Wigner}
\rho(x,t) = \frac{1}{2\pi g t}\,\sqrt{4Ngt-x^2}
\end{equation}
as in the equilibrium Dyson gas in the harmonic confining potential, $U(x)=\frac{1}{2}\epsilon x^2$. The difference is that the span now diffusively expands with time: $|x|\leq 2\sqrt{Ngt}$. This unlimited expansion eventually stops in the presence of an external confining potential. For the harmonic confining potential, the crossover time is $t_c\sim \epsilon^{-1}$.  

In the confining harmonic potential, the hydrodynamic equation reads
\begin{equation}
\label{Dyson-Hilb:H}
\partial_t \rho + \partial_x\!\left(\pi g \rho \Hil[\rho]-\epsilon \rho x\right)= 0. 
\end{equation}
The stationary density $\rho(x)$ is a solution of the integral equation 
\begin{equation}
\dashint_{-R}^R dy\,\frac{\rho(y)}{x-y} = \frac{\epsilon}{g}\,x
\end{equation}
mathematically identical to \eqref{F-eq}. Therefore, $\rho(x)$ has compact support and the density profile
\begin{equation}
\label{Wigner:H}
\rho(x) = \frac{\epsilon}{\pi g}\,\sqrt{R^2-x^2}\,, \quad R = \sqrt{\frac{4gN}{\epsilon}}
\end{equation}
is the Wigner semi-circle when $|x|<R$. The span is fixed by the normalization condition \eqref{mass}. 

The density profile \eqref{Wigner} applicable when $t\ll \epsilon^{-1}$ and the stationary density profile \eqref{Wigner:H} applicable when $t\gg \epsilon^{-1}$ suggest that generally
\begin{equation}
\label{Wigner-gen}
\rho(x,t) = \frac{2N}{\pi R(t)}\,\sqrt{1-\frac{x^2}{R^2(t)}}
\end{equation}
with amplitude ensuring the normalization \eqref{mass}.  To determine the span $R(t)$ we substitute \eqref{Wigner-gen} into \eqref{Dyson-Hilb:H} and find that \eqref{Wigner-gen} provides a consistent solution when the span satisfies 
\begin{equation}
\label{R:eq}
\frac{dR}{dt} = \frac{2gN}{R} - \epsilon R
\end{equation}
Solving \eqref{R:eq} subject to $R(0)=0$ gives 
\begin{equation}
\label{R:sol}
R = \sqrt{\frac{2gN}{\epsilon}\,\big(1-e^{-2\epsilon t}\big)}
\end{equation}
Thus, in the confining harmonic potential, the evolving density profile describing the expansion into vacuum is the Wigner semi-circle \eqref{Wigner-gen} with span given by \eqref{R:sol}.

\section{PT symmetry}
\label{app:PTsym}

Here we show that in the annealed case, the boundary-value problem, i.e., Eqs.~\eqref{MFT:qp-simple} supplemented by the appropriate boundary conditions at $t=0$ and $t=1$, is invariant under the PT symmetry
\begin{equation}
t \rightarrow 1-t  \quad  \hbox{ and } \quad  x \rightarrow -x  \, . 
\end{equation}

It proves useful to consider the auxiliary variables
\begin{equation}
  P =   \partial_x p \quad  \hbox{ and } \quad  \bar{P} = \partial_x \bar{p}
  \, , 
\end{equation}
with $\bar{p}$ defined by \eqref{eq:defpbar1}. Equation \eqref{eq:defpbar2} and \eqref{eq:qpbarp} can be rewritten as $\pi \Hil[q] =  \bar{P} - P$ and
$\pi q = 1 + \Hil[P- \bar{P}]$. Taking the Hilbert transform of \eqref{MFTq} and the space-derivative of \eqref{MFTp}, we obtain the evolution equations 
\begin{subequations}
\label{MFT-PP}
\begin{align}
  \partial_t P  &=   \partial_x\! \left(  - P \bar{P} +
   \Hil[ P] +  \Hil[ P \Hil[P- \bar{P}]] \right),
  \label{MFT-P} \\
  \partial_t  \bar{P} & =  \partial_x\! \left(    - \bar{P} P  -   \Hil[\bar{P}]
  +  \Hil[ \bar{P} \Hil[\bar{P} - P]]\right).
   \label{MFT-BarP} 
\end{align}
\end{subequations}
The boundary conditions (\ref{BCfin}) and \eqref{eq:boundnew} lead to 
\begin{equation}
  P(x,1) =   \lambda \delta(x)\quad
  \hbox{ and } \quad  \bar{P}(x,0) =  \lambda \delta(x)  
  \label{BC-PbarP}  
\end{equation}
for $P$ and $\bar{P}$. One can now verify that if $P$ and $\bar{P}$ satisfy \eqref{MFT-PP} and the boundary conditions \eqref{BC-PbarP}, the functions $\Phi(x,t) = \bar{P}(-x,1-t)$ and $\bar{\Phi}(x,t) =P(-x,1-t)$ also satisfy the same boundary-value problem. (The definitions of $\Phi$ and $\bar{\Phi}$ give $ \Hil[\Phi](x,t) = - \Hil[\bar{P}](-x,1-t)$ and $\Hil[\bar{\Phi}](x,t) = - \Hil[{P}](-x,1-t)$ which we used in computations.) The uniqueness of the solution yields  
\begin{equation}
   P(x,t) =  \bar{P}(-x,1-t)  \,  .
\end{equation}
Integrating over $x$, we deduce the announced PT symmetry relation \eqref{eq:sympbarp}.

\section{Moments and cumulants of the current}
\label{app:cumulants}

Keeping only the terms up to $\tau^{-1}$, we re-write \eqref{pdf-Expr2} as 
\begin{equation}
\label{Prob:J}
{\rm Prob}(J) \simeq {\mathcal N} e^{-bJ^2}\left[1+bJ^2\,\frac{\log(\tau/J)+\frac{1}{2}}{\tau}\right].
\end{equation}
The normalization factor ${\mathcal N}$ can be readily computed, but the following computation of the leading asymptotic behavior of the moments and cumulants of the current is organized in such a way that we do not need an explicit expression for ${\mathcal N}$. Hereinafter we use the shorthand notation $b=\frac{\pi^2 \beta}{4\tau}$, and we remind that $\tau=\log(2\pi T)$. We define 
\begin{equation}
\label{Im:def}
  I_m =  2 \int_0^\infty dJ\, J^{2m}  e^{-bJ^2}\left[1+bJ^2\,\frac{\log(\tau/J)+\frac{1}{2}}{\tau}\right]
\end{equation}
or equivalently
\begin{equation*}
  I_m =  \int_0^\infty \frac{du}{u}\left(\frac{u}{b}\right)^{m+\frac{1}{2}}  e^{-u}\left[1+u\,\frac{\log\tau-\log u + B}{2\tau}\right]
\end{equation*}
with $B=1+\log(\pi^2 \beta/4)$. The moments of the current are the ratios
\begin{equation*}
 \langle J^{2m} \rangle  =  \frac{I_m}{I_0}\,.
\end{equation*}
In the following calculations we use the digamma function defined via $\psi(z)=\Gamma'(z)/\Gamma(z)$. The digamma function admits the integral representation \cite{Abram}
\begin{eqnarray}
        \Gamma(z)\psi(z) = \int_0^\infty \frac{dy}{y}\,e^{-y} y^{z-1}\,  \log y  
\end{eqnarray}
and satisfies $\psi(z +1) = \psi(z) + 1/z$ following from the difference equation $\Gamma(z+1)=z\Gamma(z)$ for the gamma function. At half-integer values, the digamma function simplifies to \cite{Abram}
\begin{equation}
  \psi(m + \tfrac{1}{2}) = -\gamma - 2 \log 2 +
  \sum_{k=1}^m \frac{2}{2 k -1 }  
\end{equation}
where $\gamma=0.57721566\ldots$ is the Euler constant.  With the help of these formulas we obtain
\begin{eqnarray}
\label{J-m}
  &&  \langle J^{2m} \rangle  =
  \frac{\Gamma\big(m + \frac{1}{2}\big)}{b^m\,\Gamma\big(\frac{1}{2}\big)}    \\ 
  &&\times 
  \left[1 + m \frac{\log \tau + B-\psi\big(\frac{3}{2}\big)}{2\tau}-\frac{(2m+1)u_m}{4\tau}\right] \nonumber 
\end{eqnarray}
where $u_{m} = \sum_{1\leq k \leq m}  \frac{2}{2 k +1}$. Defining $\sigma$ via
\begin{equation}
 \sigma^2 = b^{-1} \left[1 + \frac{\log \tau + B-\psi\big(\frac{3}{2}\big)}{2\tau}\right]
\end{equation}
we re-write \eqref{J-m} as
\begin{equation}
\label{J-m-Gauss}
 \langle J^{2m} \rangle  =
  \frac{\Gamma\big(m + \frac{1}{2}\big)}{\Gamma\big(\frac{1}{2}\big)} \,\sigma^{2m}\left[1-\frac{(2m+1)u_m}{4\tau}\right] 
\end{equation}
in the dominant order in $\tau^{-1}$. The last factor represents the correction to the Gaussian behavior. If it were absent, the current would be a Gaussian random variable with vanishing cumulants, apart from the second cumulant, the variance. The correction to Gaussianity can be evaluated by computing the ratio
\begin{eqnarray}
  \frac{\langle J^{2m+2} \rangle }{(2m +1) \langle J^{2} \rangle  \langle J^{2m} \rangle} = 1 - \frac{u_m}{2 \tau}
  \nonumber 
\end{eqnarray}
Similar results  \cite{foglerProbability1995,DysonBOOK} describe the statistics of the number of eigenvalues of a random $N\times N$ matrix in an interval containing many eigenvalues on average, but still small compared to $N$.

Finally, we convert moments into cumulants.  The exponential moment generating function is 
\begin{equation}
\label{eqappM}
\begin{split}
 M(z) & =   1 + \sum_{m \ge 1} \frac{z^{2m}}{ (2m)!} \langle J^{2m} \rangle
 = {\rm e}^{\sigma^2 z^2/4}  - \frac{U(z)}{4\tau} \\
 U(z) & =  \sum_{m \ge 1} \frac{(\sigma z)^{2m}}{ (2m)!}  \frac{\Gamma\big(m + \frac{1}{2}\big)}{\Gamma\big(\frac{1}{2}\big)} \,(2m+1)u_m
\end{split}
\end{equation}
With the help of the identity
\begin{eqnarray}
  {\rm e}^{-z} \sum_{m \ge 1} \frac{\big(m + \frac{1}{2}\big)
    u_{m} z^m}{m!}    =  z  +  \sum_{m \geq 2}
     \frac{(m-2)! (-4z)^m}{2\, (2m)!}\nonumber  
\end{eqnarray}
we rewrite the moment generating function as 
\begin{eqnarray}
 M =  e^{\sigma^2 z^2/4}\! \left(1 - \frac{ \sigma^2 z^2 + 2\sum_{m \ge 2}
   \frac{(-\sigma^2 z^2)^m (m-2)!}{(2m)!}) } {8\tau} \right) \nonumber
\end{eqnarray}
and take the logarithm to give the announced expression \eqref{eq:formulecumulant} for the cumulants. 

The growth laws \eqref{eq:formulecumulant} are logarithmic. Therefore, unknown sub-leading terms may provide significant contributions, and numerically observing predicted growth laws may require simulations at astronomically large times.  It would be interesting to deduce the sub-leading terms for the second and fourth cumulants that exhibit the slowest growth. One could guess that the subleading contributions are finite
\begin{equation}
\label{J:24}
\begin{split}
\langle J^2\rangle_c & = \frac{1}{\pi^2 \beta}\, \log T + O(1)\\
\langle J^4\rangle_c & = -\frac{4}{\pi^4 \beta^2}\, \log T + O(1)
\end{split}
\end{equation}
However, the proliferation of the iteracted logarithms [cf. Eqs.~\eqref{Prob:J}--\eqref{Im:def}] hints for subleading terms growing as $\log(\log T)$. Deriving sub-leading contributions for the variance and the fourth cumulant is perhaps possible (albeit laborious) by refining the above analysis. 

We have not yet proved the identity that led to the neat general formula for the cumulants. We discovered it empirically, and using Mathematica, we checked the identity up to $m=1000$.

\section{Crossing of particles in Dyson gas}
\label{app:crossing}

Here we analyze the evolution of two Brownian particles with equal diffusion coefficients $D$ interacting via logarithmic potential, $V(z)=-g\log|z|$, where $z$ is the interparticle distance. This seemingly trivial system exhibits intriguing behaviors. Some of these behaviors probably appear in the literature such as \cite{RogersShi, CepaLepingle, cepa2007no}. Extracting such results is not straightforward since mathematical work aims at rigorous analyses of general and complicated systems (many particles, different diffusion coefficients, etc.) and relies on advanced probabilistic tools. Here we rely on physics tools, particularly scaling. One potential novelty is that in addition to repulsive interactions, $\beta=g/D>0$, studied in the realm of the Dyson gas, we comment on attractive interactions. 

The center mass of the system, $(x_1+x_2)/2$, undergoes Brownian motion with diffusion coefficient $D/2$. The interparticle distance $z=x_2-x_1$ exhibits more rich behavior. We interpret $z$ as the coordinate of a single particle and observe that its diffusion coefficient is $2D$, and the particle is subjected to the force $2g/z$. 

The problem of diffusing a single particle in a logarithmic potential arises in several physical problems, see e.g. \cite{Barkai10,Gunter12}. The discrete random walk version is even older problem introduced by Gillis \cite{Gillis}. According to the Gillis model, the particle at site $k\in \mathbb{Z}$ hops to $k\pm 1$ with probability $\frac{1}{2}\left(1 \pm \frac{\delta}{k}\right)$ when $k\ne 0$, and symmetrically when $k=0$. The Gillis model exhibits surprisingly rich behaviors \cite{Gaia} depending on parameter $\delta$ that plays a role similar to $\beta$ in the Dyson model. There are certainly distinctions between the Gillis model and our continuous model, mostly due to the singularity at the origin that is absent in the Gillis model. 

First we show that two particles can meet (and it happens with probability one) only when $\beta<1$; when $\beta\geq 1$, the crossing is impossible. Denote by $S(z,t)$ the probability that two particles, initially separated by distance $z>0$, never meet during the time interval $(0,t)$. This survival probability satisfies the backward Kolmogorov equation
\begin{equation}
\label{Kolmogorov}
\partial_t S(z,t) = \frac{2g}{z}\,\partial_z S(z,t) + 2D \partial_z^2 S(z,t).
\end{equation}
The initial condition is
\begin{equation}
\label{IC:S}
S(z>0, t=0) = 1.
\end{equation}
The boundary condition 
\begin{equation}
\label{BC:S}
S(z=0, t>0) = 0
\end{equation}
reflects that we seek the probability of no crossings. The initial-boundary-value problem \eqref{Kolmogorov}--\eqref{BC:S} is invariant under the one-parameter group of transformations
\begin{equation}
\label{group:S}
S\to S, \qquad z\to a z, \qquad t\to a^2 t.
\end{equation}
Therefore the solution has a self-similar form 
\begin{equation}
\label{scaling:S}
S(z,t) = \mathcal{S}(Z)\,, \qquad Z = \frac{z}{\sqrt{8 D t}}\,.
\end{equation}
Substituting the scaling ansatz \eqref{scaling:S} into the backward Kolmogorov equation \eqref{Kolmogorov} we arrive at a linear ordinary differential equation (ODE)
\begin{equation}
\label{S-eq}
\frac{d^2\mathcal{S}}{dZ^2} + \left(\frac{\beta}{Z}+2Z\right)\frac{d\mathcal{S}}{dZ} = 0
\end{equation}
which is integrated to yield 
\begin{equation}
\label{S-sol-gen}
\mathcal{S}(Z) = C_1 \int_0^Z \frac{d\zeta}{\zeta^\beta}\,e^{-\zeta^2} + C_2
\end{equation}
Using Eqs.~\eqref{IC:S}--\eqref{BC:S}, we fix the amplitudes in the general solution \eqref{S-sol-gen}. Different behaviors emerge depending on whether $\beta\geq 1$ or $\beta<1$. When $\beta\geq 1$, the solution is trivial: $C_1=0$, i.e., $S(z,t) = 1$. Thus, two Brownian particles subjected to the logarithmic potential never cross when $\beta\geq 1$. When $\beta<1$, another amplitude vanishes, $C_2=0$, and the solution is non-trivial:
\begin{equation}
\label{S-sol}
\mathcal{S}(Z) = \frac{2}{\Gamma\big(\frac{1-\beta}{2}\big)} \int_0^Z \frac{d\zeta}{\zeta^\beta}\,e^{-\zeta^2}.
\end{equation}
In the long time limit
\begin{equation}
\label{S-asymp}
S(z,t) \simeq  \frac{1}{\Gamma\big(\frac{3-\beta}{2}\big)}\, \frac{z^{1-\beta}}{(8Dt)^\frac{1-\beta}{2}}\propto t^{-\frac{1-\beta}{2}}\,.
\end{equation}
The persistence exponent $\theta$ defining the temporal decay of the survival probability, $S(z,t)\propto t^{-\theta}$ when $z=O(1)$ and $t\gg 1$, is therefore
\begin{equation}
\label{pers}
\theta = 
\begin{cases}
\frac{1-\beta}{2}  & \beta <1  \\
0                         & \beta \geq 1
\end{cases}
\end{equation}
The classical persistence exponent $\theta=\frac{1}{2}$ is recovered in the non-interacting case, $\beta=0$. Finally, we emphasize that Eq.~\eqref{S-sol} gives the survival probability not only for sufficiently weak repulsive forces, $0<\beta<1$, but also for attractive forces $\beta<0$. 

Finally, let us look at the probability density $P(z,t)$. This distribution evolves according to the Fokker-Planck equation (also known as forward Kolmogorov equation)
\begin{equation}
\label{FPE}
\partial_t P(z,t) = -\partial_z\!\left[\frac{2g}{z}\, P(z,t)\right] + 2D \partial_z^2 P(z,t)
\end{equation}

The natural initial condition, $P|_{t=0} = \delta(z-z_0)$, has an additional length scale $z_0$, and therefore the initial-value problem does not admit a scaling solution valid for all $t>0$. In the long-time limit, however, the solution approaches the scaling form 
\begin{equation}
\label{scaling:P}
P(z,t) =  \frac{1}{\sqrt{8 D t}}\,\mathcal{P}(Z)\,, \qquad Z = \frac{z}{\sqrt{8 D t}}\,.
\end{equation}
The time dependent factor in front of the scaled density $\mathcal{P}(Z)$ ensures conservation of the $z$ particle and leads to the normalization condition 
\begin{equation}
\label{norm}
\int_{-\infty}^\infty dZ\,\mathcal{P}(Z)=1. 
\end{equation}
Substituting the scaling ansatz \eqref{scaling:P} into the Fokker-Planck equation \eqref{FPE} we arrive at a linear ODE
\begin{equation}
\label{P-eq}
\frac{d^2\mathcal{P}}{dZ^2} = \frac{d}{dZ} \left(\frac{\beta}{Z}\mathcal{P}-2Z\mathcal{P}\right)
\end{equation}
which we integrate twice and obtain 
\begin{subequations}
\begin{align}
\label{P-sol}
\mathcal{P}(Z) &= \frac{2}{\Gamma\big(\frac{1+\beta}{2}\big)} Z^\beta\,e^{-Z^2}\,, \qquad \beta\geq 1 \\
\label{P-sol-attr}
\mathcal{P}(Z) &= \frac{1}{\Gamma\big(\frac{1+\beta}{2}\big)} |Z|^\beta\,e^{-Z^2}\,, \quad 0<\beta<1
\end{align}
\end{subequations}
Particles never cross when $\beta\geq 1$. Assuming for concreteness $z_0>0$, we have $\mathcal{P}(Z) = 0$ when $Z<0$. Hence, we used $\int_0^\infty dZ\,\mathcal{P}(Z)=1$ to fix the amplitude and obtained \eqref{P-sol} applicable on the half-line $Z>0$. When $\beta<1$, we used the normalization condition \eqref{norm} and obtained \eqref{P-sol-attr} now applicable on the entire line. 

Re-considering the derivation of Eq.~\eqref{P-sol-attr} one concludes that Eq.~\eqref{P-sol-attr} applies in a wider $-1<\beta<1$ range. Thus, the force can be either repulsive or attractive, but it must be sufficiently weak: $|\beta|<1$, equivalently $|g|<D$. 

Collapse is expected when $\beta\leq -1$: Particles cannot separate after crossing. Crossing necessarily happens for potentials with $\beta<1$.  If $\beta\in (-1,1)$, particles keep their individuality---after crossing, particles separate. In the long time limit, the density has the scaling form \eqref{scaling:P} with scaled density \eqref{P-sol-attr}. In the $\beta\leq -1$ range, particles never separate after collisions. The above analysis is non-rigorous, so the suggested behaviors are conjectural. 

Collapse may occur both in classical and quantum domains. A non-relativistic quantum particle of mass $m$ in an attractive $1/r^s$ potential falls into the center \cite{Landau-QM,Gupta,Essin} when $s>2$; for the inverse-square potential $V(r)=-V_0/r^2$, the collapse occurs when $V_0\geq \frac{\hbar^2}{8m}$. For the Dirac-Kepler problem, the collapse occurs \cite{Zeldovich72,Dirac-Kepler} already for sufficiently strong attractive Coulomb potential. In these examples, the collapse emerges from the inconsistent behavior of the stationary solutions of the Schr\"{o}dinger or Dirac equation. 

We treated a two-particle system where the problem reduces to the evolution of a single particle. In the Dyson gas with many particles just before the two-particle collision, we can ignore all other particles. The details of the evolution of the Dyson gas with $N$ particles and attractive logarithmic interactions with $\beta\leq -1$ are interesting even though the system's fate is trivial---all particles eventually merge. Clusters of various masses (individual particles, pairs of particles, triplets of particles, etc.) continue to merge in the pre-collapse phase. The description of such an aggregation process is an intriguing challenge. 

\bibliography{dbm}

\end{document}